# Synchrotron X-Ray Multi-Projection Imaging (XMPI) for High-Resolution 4D Characterization of Multiphase Flows

Tomas Rosén[1,*], Zisheng Yao[2,*], Jonas Tejbo[1], Patrick Wegele[1], Julia K. Rogalinski[2], Frida Nilsson[3], Kannara Mom[4], Zhe Hu[2], Samuel A. McDonald[5], Kim Nygård[5], Andrea Mazzolari[6], Alexander Groetsch[3], Korneliya Gordeyeva[1], L. Daniel Söderberg[1], Fredrik Lundell[3], Lisa Prahl Wittberg[3], Eleni Myrto Asimakopoulou[2], Pablo Villanueva-Perez[2,*]

[1]Department of Fibre and Polymer Technology, KTH Royal Institute of Technology, Stockholm, Sweden

[2]Synchrotron Radiation Research and NanoLund, Lund University, Lund, Sweden

[3]Department of Engineering Mechanics, KTH Royal Institute of Technology, Stockholm, Sweden

[4]TIMC, Université Grenoble Alpes, Grenoble, France

[5]MAX IV Laboratory, Lund University, Lund, Sweden

[6]Department of Physics and Earth Science, Ferrara University, Ferrara, Italy

*Corresponding authors: Tomas Rosén, Zisheng Yao and Pablo Villanueva-Perez

**Email:** trosen@kth.se, zisheng.yao@sljus.lu.se, pablo.villanueva_perez@sljus.lu.se

**Abstract**

Multiphase flows where particles, bubbles, or droplets are suspended in a fluid govern critical processes in biology, medicine, materials processing, and geophysics. However, observing their microscale dynamics in opaque systems has remained a fundamental challenge. We present Synchrotron X-ray Multi-Projection Imaging (XMPI), a novel approach enabling four-dimensional (3D+time) tracking of microparticles in dense suspension flows without requiring sample rotation. By capturing simultaneous projections from multiple angles using beam-split X-rays at synchrotron facilities, we resolve instantaneous particle positions and trajectories in opaque fluids such as blood. We demonstrate the potential of XMPI through individual particle tracking velocimetry (3D PTV) in dilute conditions, as well as multi-projection optical flow analysis in dense suspensions. The methodology provides otherwise inaccessible experimental validation for particle-resolved computational fluid dynamics models and allows, e.g., observation of inertial focusing effects and microstructural dynamics relevant to suspension rheology and biomedical flows. This work paves the way for high-resolution, time-resolved 4D imaging of complex multiphase flows across a range of scientific and industrial applications. Combining XMPI with recent AI-supported 4D reconstruction algorithms opens a new spatiotemporal frontier for high-speed, rotation-free microtomography.

**Nomenclature Table**

| Symbol | Description | Unit |
|---|---|---|
| $g$ | Gravitational acceleration | m/s$^2$ |
| $R_p$ | Mean radius of particles | m |
| $R$ | Inner radius of Kapton tube | m |
| $r$ | Radial position from tube center | m |
| $\Delta r$ | Particle-free layer thickness closest to tube wall | m |
| $\mu$ | Viscosity | Pa·s |
| $\rho$ | Density of SHGS | kg/m$^3$ |
| $\rho_s$ | Density of glycerol | kg/m$^3$ |
| $\rho_N$ | Number density of particles | 1/m$^3$ |
| $Q$ | Flow rate | m$^3$/s |
| $V_{sed}$ | Theoretical sedimentation velocity | m/s |
| $V_{max}$ | Theoretical maximum velocity | m/s |
| Re | Reynolds number | dimensionless |
| $x$, $y$ | Horizontal coordinates in lab-frame | m |
| $z$ | Vertical coordinate | m |
| $u_z$ | Vertical velocity | m/s |
| $\xi_1$, $\xi_2$ | Horizontal coordinates in projection 1 (P1) and projection 2 (P2) | m |
| $E$ | Energy of the X-ray beam | J |
| $\Delta E$ | Bandwidth of X-ray beam | J |

**List of Abbreviations**

| Abbreviations | Full description |
| --- | --- |
| 3D | Three dimensional (space) |
| 4D | Four-dimensional (space + time) |
| CFD | Computational Fluid Dynamics |
| ECT | Electrical Capacitance Tomography |
| MRV | Magnetic Resonance Velocimetry |
| OCT | Optical Coherence Tomography |
| OF | Optical Flow analysis |
| P1, P2 | Projection 1, Projection 2 |
| PIV | Particle Image Velocimetry |
| PTV | Particle Tracking Velocimetry |
| SHGS | Silver-coated Hollow Borosilicate Glass Spheres |
| UDV | Ultrasound Doppler Velocimetry |
| XCT | Computed Tomographic X-ray Imaging |
| XFEL | X-ray Free-Electron Laser |
| XMPI | X-ray Multi-Projection Imaging |

## 1. Introduction

Multiphase flows — involving particles, bubbles, or droplets suspended within a fluid — govern critical processes across biology, industry, and geophysics (Powell 2008; Poelma 2020). Examples span an exceptionally wide range of materials: blood, ink, paint, paper pulp, mud, lava, concrete, slurries, ketchup, and toothpaste all represent dense suspensions where microscopic particles govern macroscopic behavior. Although these flows appear continuous at macroscopic scales, they consist of interacting micrometer-sized components whose local dynamics strongly affect the bulk properties of the system. These microscale interactions influence processes such as clot formation in blood, landslides in mud, fiber alignment in papermaking, and flow resistance in dense industrial suspensions (Stickel and Powell 2005; Butler and Snook 2018; Lundell, Söderberg, and Alfredsson 2011; Ness, Seto, and Mari 2022).

Understanding the physical behavior of these complex systems requires capturing their full four-dimensional (3D + time) dynamics at micrometer resolution. The ability to resolve how particles move, collide, and arrange within dense suspensions provides critical insight for predicting effective viscosity, flow-induced microstructures, and transitions to solid-like states such as jamming or aggregation. Despite advances in theoretical modeling and computational simulations (Freund 2014; Aidun and Clausen 2010; Sommerfeld 2017), experimental measurements remain essential for validation and for providing ground truth for these models.

Obtaining such time-resolved 3D experimental information for dense suspensions remains technically demanding (Powell 2008; Poelma 2020). Optical methods, such as laser-based particle tracking or optical coherence tomography (OCT), are limited to transparent or semi-transparent systems, often requiring index-matching strategies that are not applicable to many real-world suspensions (Cheng et al. 2011; Zade et al. 2018; Byron and Variano 2013; Haavisto, Koponen, and Salmela 2014). Other approaches, such as ultrasound Doppler velocimetry (UDV), magnetic resonance velocimetry (MRV), and electrical capacitance tomography (ECT), can access opaque systems but typically yield limited spatial or temporal resolution and often only provide averaged or indirect flow information (Poelma 2017; Fukushima 1999; Ma et al. 2019; Markl, Kilner, and Ebbers 2011; Van Ooij et al. 2012; Zhang et al. 2014).

X-ray imaging offers a fundamentally different approach to overcome the optical opacity of these systems. Owing to their short wavelength and strong penetration capability, X-rays enable high-resolution imaging of opaque systems, providing access to multiphase flows at the micrometer scale (Heindel 2024; Kastengren and Powell 2014), including blood flows (Kim and Lee 2006; Lee, Ha, and Nam 2010; Irvine et al. 2008), cavitation flows (Jahangir et al. 2019), foams (Schott et al. 2023), drainage dynamics in porous media (Tekseth et al. 2024), and fluidized bed reactors (Mudde 2010). In addition, X-ray imaging is not affected by refractive distortions from curved or inclined channel walls, which commonly limit optical methods. This makes X-ray based techniques particularly robust for studies involving complex geometries or non-planar interfaces. In these studies, X-rays were generated through lab devices like tube sources and electron guns, or advanced X-ray sources such as synchrotron light sources and X-ray free-electron lasers (XFELs). Yet, capturing fully time-resolved 4D information using X-rays has remained limited.

State-of-the-art 4D X-ray approaches acquire time-resolved X-ray images (radiographs) at multiple angles by either rotating the sample, *i.e.* computed tomographic X-ray imaging (XCT) (Withers et al. 2021), or by subjecting the sample to multiple beams from different directions (Heindel, Gray, and Jensen 2008). Although time-resolved 4D XCT has been demonstrated for multiphase flows (Mäkiharju et al. 2022), the rotation limits its applicability to cases where the time scale for the flow is significantly longer than the time scale for rotation. The latter also poses a restriction since rapid rotation induces inertial forces in the rotating frame-of-reference that can lead to secondary flows. Alternatively, multiple X-ray beams can illuminate a stationary sample simultaneously and orthogonally, as demonstrated with laboratory sources (Chen, Zhong, and Heindel 2019; Bieberle and Barthel 2016; Neumann et al. 2019). For example, electron guns can generate an X-ray fan around the sample to provide a single 2D slice with kHz temporal resolution but millimeters spatial resolution (Bieberle and Barthel 2016; Neumann et al. 2019). Multiple state-of-the-art X-ray lab sources can then be mounted for stereographic particle tracking in 4D (Chen, Zhong, and Heindel 2019) to retrieve spatiotemporal resolutions on the order of 50 µm at 300 Hz (Zwanenburg, Williams, and Warnett 2021; Vavřík et al. 2017). In short, such systems are suitable for tasks such as 3D particle tracking with a field of view of ~10 centimeters and a moderate spatiotemporal requirement, but the spatiotemporal resolution they can provide is ultimately limited by the flux from the lab-based X-ray source (Heindel 2011; Aliseda and Heindel 2021; Villafañe et al. 2025). On the other hand, the high X-ray flux from synchrotron light sources or XFELs can enhance the spatiotemporal resolution to sub-micrometer resolution and kHz acquisition rates and beyond (Liang et al. 2023; Asimakopoulou et al. 2024; Villanueva-Perez et al. 2023; Olbinado and Rack 2019; Campbell et al. 2021), although generating multiple simultaneous beams for 3D imaging at these facilities remains nontrivial.

X-ray multi-projection imaging (XMPI) at synchrotron light sources and XFELs was recently demonstrated (Villanueva-Perez et al. 2018; Hoshino et al. 2011; Duarte et al. 2019), enabling 4D movies with micrometer resolution and kHz frame rates (and beyond) without rotating the sample (Liang et al. 2023; Asimakopoulou et al. 2024; Villanueva-Perez et al. 2023). This technique splits a single X-ray pulse to provide angularly separated beams, allowing single-shot 3D imaging and unparalleled 4D acquisition rates.

However, XMPI results in limited volumetric information due to the low number of projections. Thus, novel 4D reconstruction algorithms have been developed (Zhang et al. 2024; Zhang et al. 2023) that combine ideas from the following techniques: i) iterative 3D reconstruction (Herman 2009), like algebraic reconstruction, ii) a combination of different experiments and samples (Punjani and Fleet 2023), such as cryo-electron microscopy, iii) physical priors (Goodman 2005), *i.e.*, considering the interaction and propagation of X-rays with matter, and iv) deep-learning concepts to reconstruct 3D and 4D from sparse projections (Mildenhall et al. 2021), such as neural radiance fields. Combining these algorithms with XMPI opens new possibilities for studying time-resolved micron-sized features of dense particle suspension flows at kHz acquisition rates (and beyond). It thus enables three-dimensional studies of high-speed and turbulent flows while eliminating experimental artifacts.

Here, we demonstrate how XMPI enables direct, rotation-free 4D particle tracking in both dilute and concentrated suspensions, including complex fluids such as human blood. We show that individual microparticles can be tracked in real time, and that statistical flow properties such as velocity profiles and migration phenomena can be quantified even in dense systems. This approach opens new experimental possibilities for studying the microscale dynamics of opaque multiphase flows across a wide range of scientific and technological domains.

## 2. Method

### *2.1 Overview of experiment*

The XMPI experiment was performed at the ForMAX beamline, MAX IV, Lund, Sweden. Photos of the experimental setup are provided in Fig. 5 in Appendix A. The direct beam was split using the Bragg reflection of Silicon and Germanium crystals (**Fig. 1**). The X-ray photon energy was 16.55 keV, with two beamlets at an angle of approximately 48°, simultaneously illuminating the sample. The transmitted intensity from each beam was recorded by two X-ray microscopes with an effective pixel size of 1.3 µm, positioned several centimeters after the sample. This propagation distance can help to observe low-contrast features due to phase contrast (Snigirev et al. 1995; Suzuki, Yagi, and Uesugi 2002). Note that the difference in propagation distances between the sample and the two microscopes leads to a visible difference in propagation effects in the two projections. The sample consists of a Kapton tube with an inner diameter of $2R$ = 0.72 mm with a flow driven by a syringe pump. To highlight this technique's potential, we demonstrate two multiphase flow scenarios, with movies provided as Supplementary Material:

- Suspended microparticles in glycerol.
- Suspended microparticles in human blood.

### *2.2 Sample*

The particles used in this work were silver-coated hollow borosilicate glass spheres (SHGS, Dantec Dynamics) with mean diameter $2R_p$ = *10 µ*m and density $\rho$ = 1400 kg/m$^3$. As a suspension media, we used glycerol with a density of $\rho_s$ = 1260 kg/m$^3$ and viscosity $\mu$ = 1.4 Pa·s.

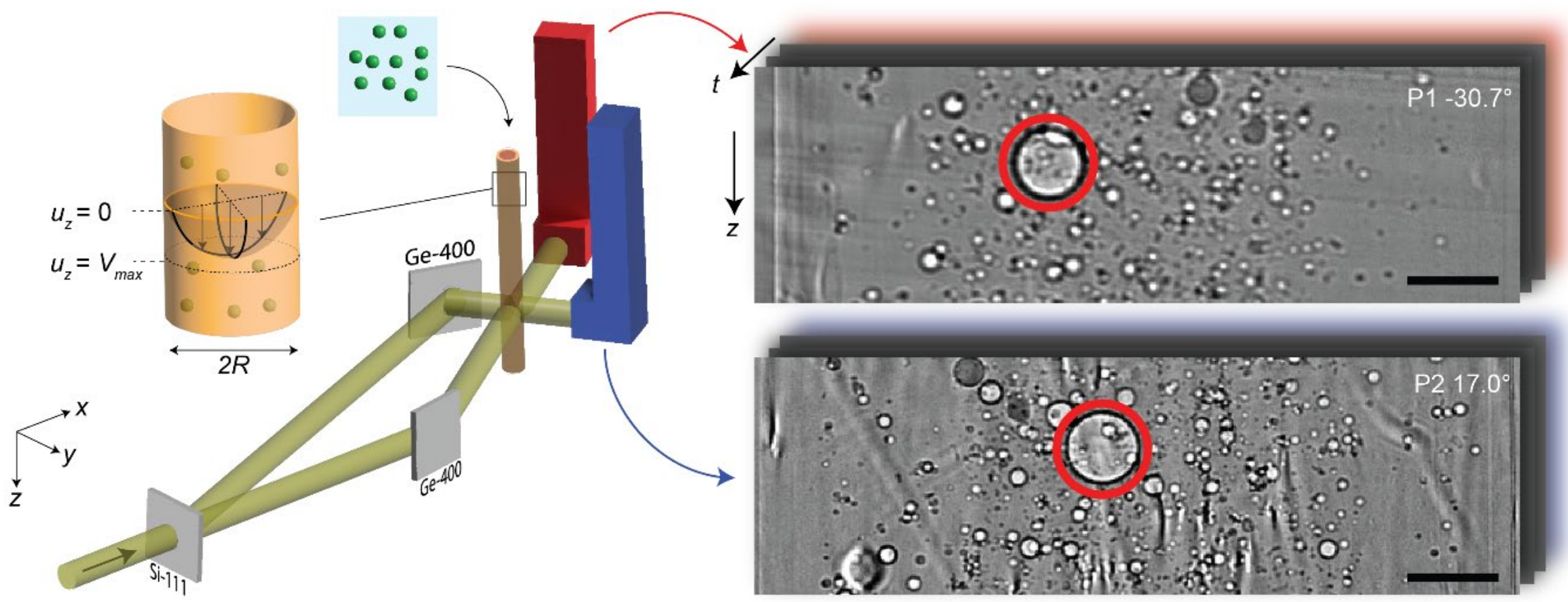

**Fig. 1** Experimental setup and XMPI concept for multiphase flow experiments. The direct beam is split and recombined to produce two stereographic projections, enabling 4D tracking of individual particles from two directions. The red circle highlights a large particle seen in both directions. In two projections, different performances of edge enhancement of particles are observed due to slightly different sample-detector distances. Scale bars: 100 µm. Angles related to the direct beam direction. See Movie S1 for the recorded movie.

Human whole blood of about 40% hematocrit up for destruction, *e.g.* expired, was used for the experiments. The blood was obtained from the Department of Transfusion Medicine at Karolinska University Hospital, Huddinge, Sweden.

*2.3 Flow*

The flow was driven with a syringe pump (New Era Pump Systems NE-4000) with a 1 mL syringe at a set flow rate of $Q$ = 0.1 mL/h through a Kapton tube (Allectra 312-KAP-TUBE-07-300), with a reported inner diameter of $2R$ = 0.72 mm and 25 µm wall thickness, leading to a theoretical maximum velocity of $V_{\max} = 2Q/(\pi R^2)$ = 0.137 mm/s. The distance from the flow entering the Kapton tube to the measurement section is approximately 10 mm.

The theoretical sedimentation velocity is $V_{sed} = 2(\rho\text{-}\rho_s)gR_p^2/(9\mu) \approx 5.45\times10^{-6}$ mm/s, and since $V_{sed} \ll V_{max}$, we can assume particles at a radial position r from the center of the capillary are following the analytical Poiseuille velocity profile $u_z = (2Q/(\pi R^2))(1\text{-}r^2/R^2)$. This equation is used to fit the experimental data, using $Q$ and $R$ as fitting parameters, with residuals defined by the error in radial position r for a certain measured velocity $u_z$.

From the experimental data in this work, we found that the actual inner diameter was $2R$ = 0.78 mm and the actual flow rate $Q$ = 0.115 mL/h, leading to a maximum velocity of $V_{\max}$ = 0.134 mm/s. The Reynolds number of this flow is $Re = \rho_s \cdot (V_{max}/2) \cdot 2R/\mu = 4.7\times10^{-5}$.

*2.4 X-ray Multi-projection Setup*

The experiments were performed at the ForMAX beamline at MAX IV (Nygård et al. 2024; Tavares et al. 2014), the first operational diffraction-limited storage ring. The ForMAX beamline is in the MAX IV 3 GeV storage ring, utilizing undulators as source. At this beamline, we performed the experiments at a photon energy of 16.55 keV, which delivers one of the highest photon fluxes over a narrow bandwidth ($\Delta E/E \approx 0.01$, provided by the multilayer monochromator) and an illumination size of 1x1 mm$^2$ with a collimated beam (Yao et al. 2024). Using the direct beam, we used the crystal configuration shown in Fig. 1, inspired by Ref. (Mokso and Oberta 2015) to generate the two beamlets to perform Particle Tracking Velocimetry (PTV). The first beamlet was split from the incoming beam by using a Si-111 crystal that generated a deflection angle with respect to the main beam of –13.72 degrees. This beam was recombined into the sample position by using a Ge-400 crystal, which generated a recombination angle of 16.99 degrees with respect to the main beam (P2 in Fig. 1). The second beamlet was generated by using a Ge-400 crystal, which generated a beam deflection of –30.70 degrees (P1 in Fig. 1). Thus, the angle between the two beamlets used for PTV was 47.69 degrees. It is important to note that the selection of such angles is based on a compromise between the diffraction efficiency (1-2%) and the angular separation, ensuring both proper angular range and image quality provided by the XMPI setup (Bellucci et al. 2024; Rogalinski et al. 2025). The sample was positioned at the intersection point between the two beamlets. The two beamlets were detected with two identical indirect X-ray microscopes. The X-ray microscopes used a LuAG:Ce scintillator and a high-NA 5X magnification lens to provide an effective pixel size of 1.3 µm when coupled to an Andor Zyla 5.5 sCMOS camera. The setup was operated at 40 Hz, and the exposure time was set to the inverse frame rate to maximize the contrast-to-noise ratio. Given this configuration and experimental setup, we find that blurring of particle edges starts occurring when the particles move more than ~2 pixels during exposure, i.e., around 0.1 mm/s. To avoid blurring at this frame rate, the exposure time can be reduced at the cost of decreasing the contrast-to-noise ratio, with the field-of-view limiting the maximum velocity. Assuming a maximum allowed displacement of half the field-of-view (~500 µm) to identify particles in two consecutive frames, the maximum velocity that can be tracked is 20 mm/s in the current setup. Faster flows can be studied with kHz and beyond cameras, for example, using Photron Nova S16 (16 kHz), as previously used (Asimakopoulou et al. 2024). Such a setup has the potential to probe particle velocities of 8000 mm/s, and thus providing a theoretical possibility to study Reynolds numbers of 4000 with this change of cameras for a suspension in water.

The recorded images by both cameras were corrected to eliminate fixed pattern noise due to the illumination and the detector's dark current. For that, we acquired 100 glycerol-only images (by applying a flow of glycerol with a high flow rate to avoid residual particles in the field of view) and 100 dark-current images (images without illumination) for every experiment. The corresponding averaged glycerol-only images and averaged dark-field images were used for correction, analogous to conventional flat-field correction (Van Nieuwenhove et al. 2015).

### *2.5 Microparticle tracking of dilute suspensions*

The experiment consisted of 8000 frames simultaneously recorded on each of the two detectors at 40 Hz. The flow-coordinate $z$ was set by tracking one reference particle in each view since the $z$-coordinate was equivalent in both views. The tracking of particles was done through two preprocessing steps using Python 3.8.8 and several main steps using MyPTV (Shnapp 2022), an open-source Python-based software to track the positions of particles in 3D using the nearest neighbor method (Ouellette, Xu, and Bodenschatz 2006). Based on the particle tracking results, we get the 3D trajectory of each particle as well as their velocity. Technical details are provided in Appendix B.

### *2.6 Optical flow (OF) analysis of dense suspensions*

Optical flow (OF) analysis was performed on a 5-second sequence of corrected images (200 images) of SHGS at 10 wt.% using opticalFlowRAFT in MATLAB R2024b (MathWorks 2024). This function estimates the velocities of each pixel between video frames using the recurrent all-pairs field transforms (RAFT) algorithm (Teed and Deng 2020). The velocities were scaled using the known pixel size and frame rate, then averaged over time and vertical direction.

In the middle of the projected channel, it is important to note that particles are both slow-moving close to the walls and fast-moving on the centerline. The resulting curves thus represent the average velocity in the viewing direction. For comparison with the 3D PTV data, $10^7$ artificial particles were randomly sampled within a circle with a diameter of $2R$ and assigned a velocity according to the fitted data of $Q$ and $R$ from the 3D PTV results. Particles were binned according to their $x$-position, and an average was taken for every bin to provide a comparison to the OF results. To account for a particle-free layer with thickness $\Delta r$, artificial particles were instead sampled within a circle of radius $R$-$\Delta r$. Further technical details are provided in Appendix F.

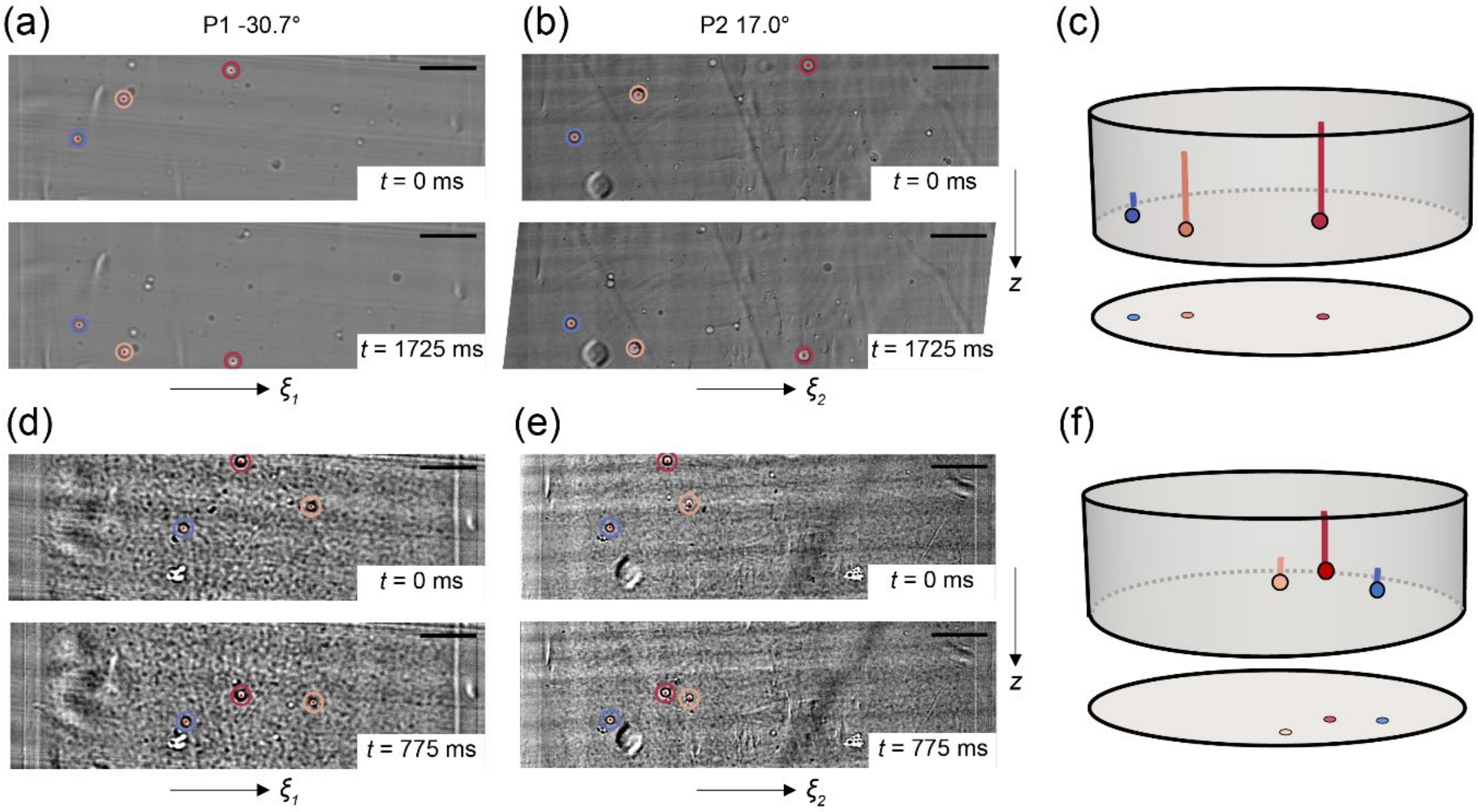

**Fig. 2** 4D particle tracking of microparticles. Three SHGS particles (colored rings) flowing in glycerol (a,b) or blood (d,e) were identified in two projections at two-time instances with red dot indicating the tracked particle center position. Reconstructed 4D trajectories between these two-time points for glycerol (c) and blood (f). Scale bars: 100 µm. Datasets for (a-b) and (d-e) are in Movie S2 and Movie S3, respectively.

## 3. Results

### *3.1 4D tracking of individual microparticles flowing in glycerol*

Firstly, we studied a laminar flow of polydisperse silver-coated hollow-glass spheres (SHGS) in glycerol (diameter between 5 and 35 µm) with a mean diameter of 10 µm. Fig. 1 illustrates a sequence of frames at a concentration of 1 wt.%. These particles, commonly used as tracers in fluid dynamics research, are well-suited for X-ray imaging (Parker and Mäkiharju 2022). Due to their hollow nature, they appear with lower attenuation than the surrounding fluid and an interface enhanced by phase-contrast imaging. At low concentrations, each particle was distinguishable in both projections by vertical coordinate $z$ and velocity. Tracking can be further verified by the polydispersity of particles, as they are distinguishable by size. **Fig. 2** shows the 3D position of each particle for a given time determined by the vertical coordinate $z$, horizontal coordinates $\xi_1$ and $\xi_2$ in each projection, and the angle between viewing directions. For example, three different particles are tracked at two different times in Fig. 2(a)-(b). The tracking was performed using a calibrated coordinate system and the MyPTV library (Shnapp 2022). By tracking the particles over time in 3D, we resolved their 4D trajectories, as depicted in Fig. 2(c). The Appendix C includes a detailed trajectory analysis in Fig. 6(a)-(b).

### *3.2 4D tracking of individual microparticles flowing in human blood*

As a second case, we studied the motion of SHGS particles in human blood as an example of a highly dense suspension. Although blood is opaque to visible light, the microparticles were easily distinguishable in our experiment. Using the aforementioned procedure, we identified particles in two projections, Fig. 2(d) and (e), allowing us to retrieve their 3D position and track them over time. The 4D trajectories of three particles in blood are depicted in Fig. 2(f), with a detailed trajectory analysis in Fig. 6(c)-(d). Although red blood cells (RBCs) were not directly visible, a speckle pattern in both projections, see Fig. 2(d) and (e), was observed. Such a pattern has been previously used to track blood flow with X-rays in individual 2D projections (Kim and Lee 2006; Lee, Ha, and Nam 2010; Irvine et al. 2008).

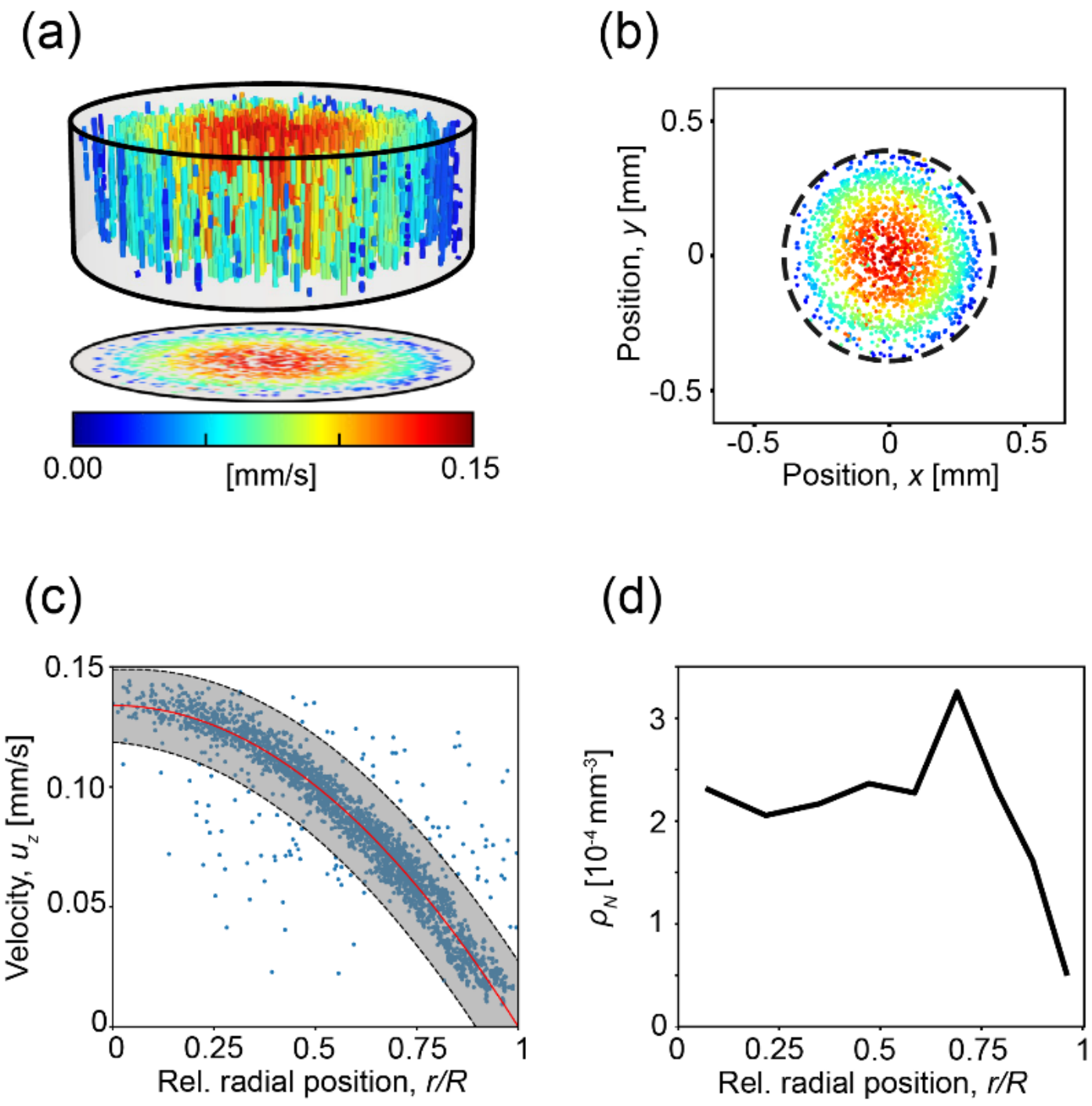

**Fig. 3** Statistical analysis of dilute SHGS particles in glycerol using 3D PTV. (a) Trajectories of 2471 tracked particles at 0.1 wt.% concentration, hue indicates vertical velocity $u_z$. (b) Horizontal distribution of particles and their velocities $u_z$ using the same hue scale. (c) $u_z$ as a function of radial position $r$ compared with theoretical cylindrical Poiseuille flow profile (black curve). (d) Particle number concentration $\rho_N$ as a function of radial position. Datasets for PTV are provided in Movie S4. A rendering of the 4D data set in (a)-(c) is provided in Movie S6.

### *3.3 Retrieving statistical properties in dilute and concentrated suspensions*

These experiments demonstrate the ability to track individual micrometer-sized particles in 4D within opaque multiphase flows using XMPI, enabling the study of stochastic phenomena not captured by techniques like OCT or MRV. Although this method can be used to track individual particles, it can also be used to study statistical properties. For instance, as small particles primarily follow the fluid motion, we can perform statistical 3D particle tracking velocimetry (3D PTV) of these trajectories, providing velocity fields comparable to other methodologies.

*3.3.1 Particle tracking velocimetry (PTV) in dilute suspensions.* To demonstrate the ability to retrieve statistical properties and validate the XMPI approach, we tracked more than 2400 SHGS particles in glycerol flowing through the cylindrical capillary, **Fig. 3**(a), determining their average horizontal positions and downstream velocity as a function of radial position, Figs. 3(b) and (c), respectively. In Fig. 3(c), a 3σ error band was estimated by considering both uncertainties of radial position (6.1 μm) and velocity (0.005 mm/s), where details can be found in the Appendix D and Fig. 7. As a result, 95.8% of all tracked particles were inside the error band. Both radial velocity and concentration profiles were extracted from low-density suspensions (0.1 wt.%), as higher concentrations complicate such studies due to challenges in detecting particle centers when multiple particles are in contact within the camera view. The figure also shows that the resulting particle velocities fit accurately to the expected laminar Poiseuille flow profile. Plotting the number density of particles $\rho_N$ versus radial position as seen in Fig. 3(d), there are clearly fewer particles in the direct vicinity of the walls. Although this is expected owing to the Segré-Silberberg effect (Segre and Silberberg 1962; Segre and Silberberg 1961; Di Carlo et al. 2007), it is unlikely that such “inertial focusing” is visible at these creeping flow conditions, and the particle-free layer likely stems from non-ideal upstream conditions and potential fouling of the tube by near-wall particles. It is still worth emphasizing the possibility of studying general migration phenomena, including inertial focusing, using XMPI and 3D PTV in opaque systems. Further 3D PTV studies with different concentrations and flow rates are in Figs. 8 and 9 of Appendix E.

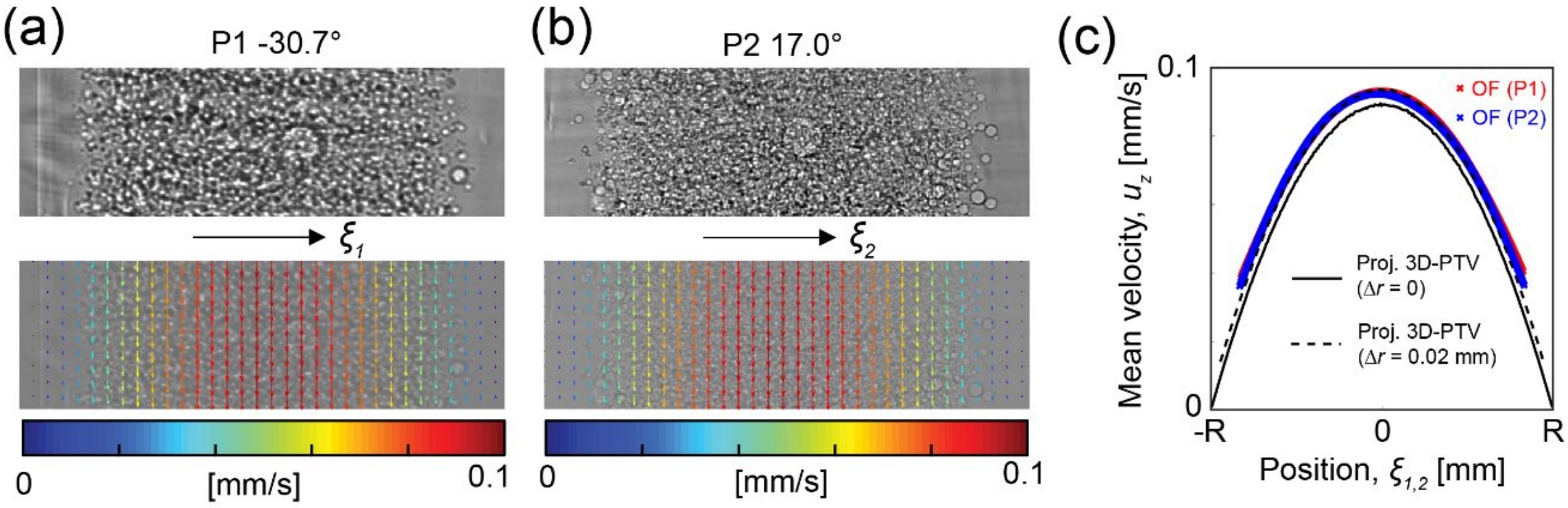

**Fig. 4** Statistical analysis of dense SHGS particles in glycerol using Optical Flow (OF) analysis. (a), (b) Projected OF at 10 wt.% for each projection. (c) Comparison of OF results (blue and red crosses) with the projected 3D PTV velocity profile from Fig. 3(c) assuming a 0.02 mm particle-free layer (black dashed curve) as well as the profile assuming no such layer. Datasets for OF analysis are provided in Movie S5.

*3.3.2 Optical flow (OF) analysis in dense suspensions.* Tracking individual particles in dense systems from two projections is very challenging, unless the particle can be distinguished through unique spatiotemporal features (*e.g.* the big particle visible in **Fig. 4**(a) and (b)). However, the optical flow methodology (OF) can be applied to obtain projected velocity profiles, as seen in Fig. 4(c), which shows the average velocity profiles from both projections for a 10 wt.% concentration. At this concentration, suspension viscosity can be assumed to be shear-rate independent. Thus, the OF velocity profile should match the averaged 3D PTV profile projected along the viewing direction. Accounting for a 0.02 mm particle-free layer, we found good agreement between the OF analysis of the dense suspension and the averaged Poiseuille profile shown in Fig. 3(c). The fact that the projected profiles from the two measurements are closely matching further confirms the cylindrical symmetry of the flow. More details about the procedure are found in Figs. 10-13 in Appendix F.

The images were also analyzed using particle image velocimetry (PIV) through PIVlab in MATLAB R2024b (Thielicke and Sonntag 2021) and a comparison is provided in Fig. 14 in Appendix F. However, given the large size of the tracers, the subpixel accuracy could cause large displacement errors. Furthermore, it is important to note that PIV is normally applied to 2D planes of illuminated particles in a laser sheet perpendicular to the viewing direction. Nevertheless, PIV analysis of integrated X-ray images from multiple views have already been used to determine the 3D-flow field (Dubsky et al. 2010), and would be an interesting topic for further research using XMPI.

## 4. Discussion and Conclusions

We demonstrated XMPI's ability to be used for multiphase flow research by tracking micron-sized particles or other microscopic features in 4D in flows without sample rotation. The method can furthermore be expanded to studies of flows through complex geometries, *e.g.,* porous media or even custom-made micro-3D-printed geometries. This method provides otherwise inaccessible experimental validation for particle-resolved computational fluid dynamics (CFD) simulations. With more projections, XMPI might be used to capture the instantaneous full 3D-velocity field of dense suspensions with 3D PIV, as described by Dubsky et al. (2010) This could be done directly on the speckle pattern from blood (Kim and Lee 2006; Irvine et al. 2008; Lee, Ha, and Nam 2010), allowing a detailed comparison of particle trajectories and simultaneous velocity fields of RBCs in 4D. In dense suspensions of solid particles, XMPI can reveal microstructural dynamics leading to aggregation, yielding, and jamming (Ness, Seto, and Mari 2022; Stickel and Powell 2005), crucial for understanding the suspension rheology. As we emphasize the experimental modality here using easily accessible open-source packages for particle tracking and OF, there can also be a further improvement to the analysis of the acquired images allowing for 4D particle tracking at concentrations above dilute conditions.

Despite the potential of micrometer-resolved XMPI for multiphase flows at synchrotron light sources, there are limitations worth mentioning. The limited field of view (currently < 10 mm), limited angular separation (currently 47.69°), and the availability of these facilities render many experiments more suitable for lab sources, especially when high spatiotemporal resolution (micrometer resolution with ~100 Hz or higher frame rates) is unnecessary. Although the temporal resolution in this experiment was modest (40 Hz), higher frame rates are achievable using faster X-ray detection schemes. The ultimate achievable spatiotemporal resolution depends on flux density, *i.e.*, the number of photons per unit of time and area on each beam. Diffraction-limited storage rings (*e.g.,* MAX IV (Tavares et al. 2014), ESRF-EBS (Raimondi 2016), and APS-U (Kerby 2023)) and X-ray free-electron lasers (*e.g.,* EuXFEL (Decking et al. 2020), LCLS (Emma et al. 2010)) increase the flux density by more than one order of magnitude compared to the previous generation of synchrotron light sources, enabling micrometer resolution with kHz framerates and beyond (Asimakopoulou et al. 2024; Villanueva-Perez et al. 2023). Higher flux also offers opportunities for more beamlets and, thus, higher angular resolution. Combining XMPI and 4D XCT reconstructions from sparse projections using machine learning (Zhang et al. 2024) suggests a future where high-speed, high-resolution imaging of multiphase flows without sample rotation becomes feasible.

To summarize, we report the first experiment of a multiphase flow using XMPI at a synchrotron light source, achieving time-resolved micrometer-scale 4D imaging of opaque fluids, such as blood. We demonstrated the possibility of tracking individual particles in 4D and the possibility of determining statistical flow properties through 3D PTV such as the velocity profile and migration behavior. We also quantified instantaneous projected velocities in dense suspensions using OF. Thus, we envision that this methodology will open up a new frontier in experimental studies of multiphase flows, with applications not limited to particle suspensions but also extending to gas-liquid flows, *e.g.,* foams, creams, and immiscible liquid-liquid flows, *e.g.,* emulsions.

## Appendix A Photos of experimental setup

Photos of the XMPI experimental setup at the ForMAX beamline at MAX IV, are provided in Fig. 5.

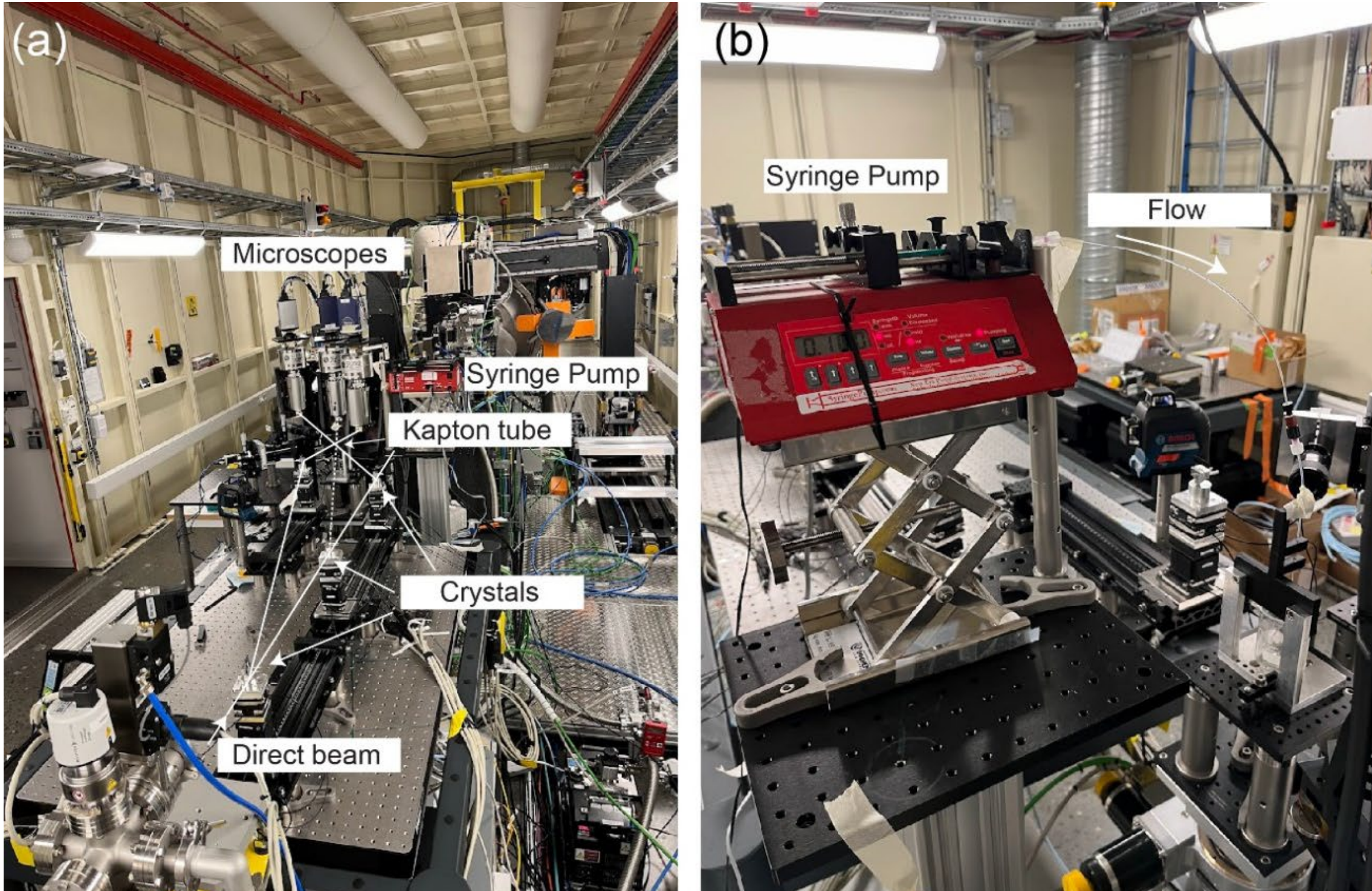

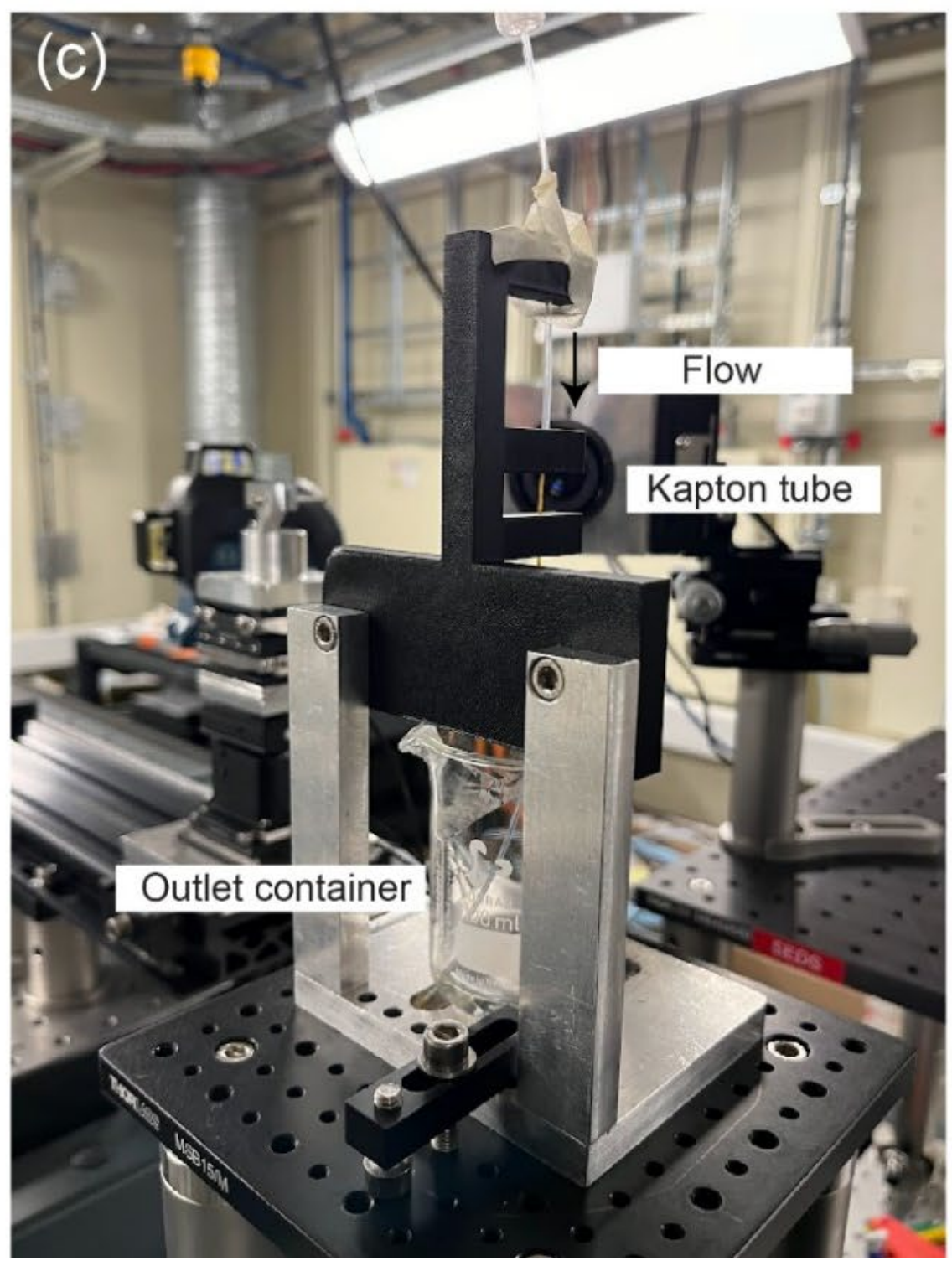

**Fig. 5** Photos of the experimental setup for XMPI of multiphase flows at the ForMAX beamline, MAX IV, Sweden. (a) XMPI setup; Note that a third crystal and microscope were mounted but not used in this study. (b) Flow setup. (c) Close-up of the measurement section.

## Appendix B Details of MyPTV-based microparticle tracking

The microparticle tracking procedure includes two preprocessing steps and several main steps using MyPTV toolbox.

In the first preprocessing step, we generated calibration images and the calibration target file for the initial camera calibration implemented later in MyPTV. First, we manually tracked a particle in both views and selected ten frames with a constant stride when this particle was within the field of view of both cameras. Second, we generated the calibration image for both views by summing up these ten frames so that ten different locations of the particle were clearly seen in both views. Third, we created the calibration target file by marking the *z* positions of the particle. The horizontal position ***r*** = (*x,y*) was set as (0, 0) temporarily for this step and will be modified later based on the tracking results.

In the second preprocessing step, we created a binary mask of moving objects to help MyPTV detect the particles more efficiently. The mask was created by thresholding pixels that differ by a certain percentage from the mean image taken of the entire set of flat-field corrected frames (5% and 7.5%, respectively, for the two detectors). Some problematic regions of the images with fluctuating pixel intensities were manually determined and removed from the binary mask.

The main steps using MyPTV are described as follows:

1. Initial camera calibration

The pinhole camera model used in MyPTV toolbox is depicted in Eq. (1), where $\vec{r}$ and $\vec{O}$ denote the particle position and camera's center position in lab space coordinates, respectively; $\eta$ and $\zeta$ denote the particle position in 2D image space coordinates; $x_h$ and $y_h$ denote the displacement to camera's center position; $\vec{e}(\eta,\zeta)$ denotes a nonlinear correction term, which is assumed to be a quadratic polynomial of $\eta$ and $\zeta$; $[R(\theta_1,\theta_2,\theta_3)]$ is the rotation matrix, which is a function of three orientational angles in 3D space.

$$\vec{r} - \vec{O} = \left( \begin{bmatrix} \eta + x_h \\ \zeta + y_h \\ |\vec{O}| \end{bmatrix} + \vec{e}(\eta,\zeta) \right) \cdot [R(\theta_1,\theta_2,\theta_3)] \quad (1)$$

In the initial camera calibration, the calibration images generated in the first preprocessing step are used as input. Although ten particle positions are generally not enough for a precise calibration, we can still get a reasonable initial camera model because the angles of two detectors in the XMPI setup are precisely known due to the diffraction condition of the used splitters. This step is done once we get the two camera models that satisfy the angular configuration of the XMPI setup with an averaged calibration error of less than 1 pixel for all ten particles shown in the calibration images. It is worth mentioning that such initial camera models do not contain the nonlinear correction term $\vec{e}(\eta,\zeta)$, which will be modified during step 5.

2. Particle segmentation

The masked images from the second preprocessing step are used as the input of this step. We use a local mean subtraction filter, a median noise removal filter, and a Gaussian blur filter provided in MyPTV to boost the performance of particle segmentation. The output of this step is the centroids of each particle with a diameter between 3 pixels and 20 pixels in both cameras.

3. Particle matching

Segmentation results are used to triangulate the 3D positions of the particles via the "particle marching" algorithm supported in MyPTV. At this step, particles are matched in both cameras so that the real positions can be derived with epipolar geometry. The most important parameter at this step is the maximum allowed triangulation error in lab-space coordinates. Practically, we set this parameter to the length of 3~5 pixels.

4. Particle tracking

The 3D trajectories of the particles are formed by linking particles from step 3 in the time domain. At this step, we use the nearest neighbor method (Ouellette, Xu, and Bodenschatz 2006) provided in MyPTV.

5. Calibration with reliably tracked particles

Long trajectories from step 4 are used to refine the camera models for each view. Practically, we select trajectories longer than 50 frames at this step. The output of this step is the updated camera models for both cameras including the nonlinear correction term $\vec{e}(\eta, \zeta)$.

It is important to note that steps 3, 4, and 5 follow an iterative process. We proceed to step 6 only when the number of trajectories and the average temporal trajectory length reach a certain level after step 4. For example, in the experiment shown in Fig. 3(a)-(c), the levels are set as 40000 total trajectories and an averaged temporal trajectory length of 10 frames.

6. Trajectory smoothing

Based on the trajectories from step 4, linear fitting is implemented to calculate the velocity $u_z$. Trajectories shorter than a threshold (typically 10-30 frames) are discarded. In the experiment shown in Fig. 3(a)-(c), the threshold is set as 20 frames.

## Appendix C Individual behavior of tracked particles

Based on the 4D trajectories shown in Fig. 2 (main text), we can analyze the trajectories in more detail both vertically and radially. Fig. 6 shows the relationship between the vertical position and the time, as well as the radial position and the vertical position, respectively, where the colors of the curves are consistent with the corresponding particles shown in Fig. 2(c) and (f) (main text). Regarding three tracked SHGS particles in glycerol, Fig. 6(a) gives both detected particle positions (markers) and the positions in linear fitted trajectories (solid line), where linearity can be clearly observed. Fig. 6(b) shows that the vertical velocities and the radial positions of these three particles are independent of their vertical positions. Regarding three tracked SHGS particles in the blood, they also keep a constant vertical velocity, as linearity can also be clearly observed in Fig. 6(c). Fig. 6(d) shows that particle 1 moves towards the center, while particle 3 moves towards the wall, which indicate a migration behavior.

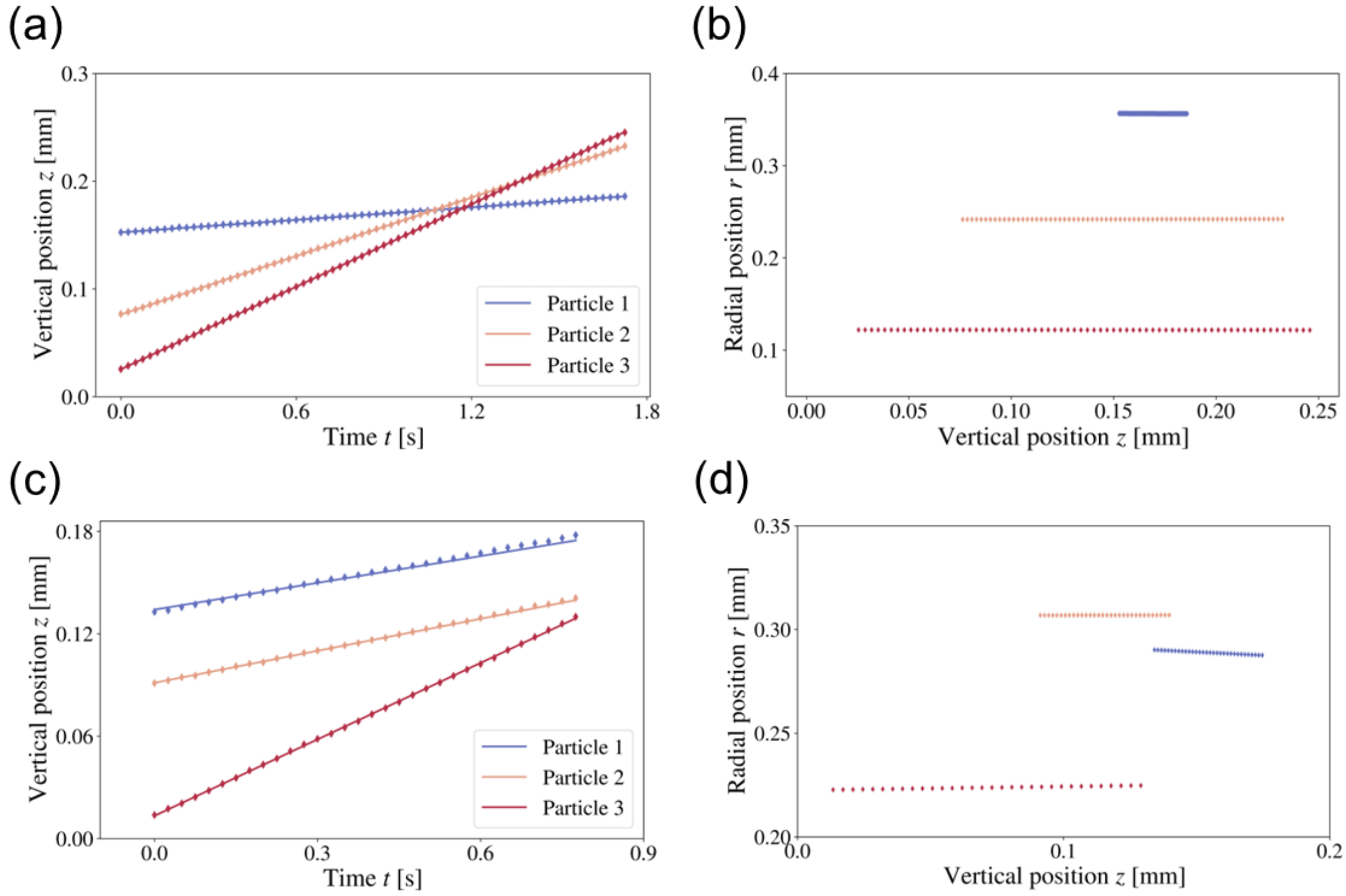

**Fig. 6** Example of the analysis of three tracked SHGS particles in glycerol (a)(b) and in blood (c)(d), respectively. (a) and (c) show how the vertical position changes with time; (b) and (d) show the relationship between the radial position and the vertical position.

**Appendix D Error analysis of tracked particles**

To calculate the uncertainty of the position of a tracked particle, the coordination systems, viewed from the direction perpendicular to the optical plane, are defined as shown in Fig. 7(a). The red and the green arrow denote the direction of two X-ray beamlets generated by our XMPI setup, respectively. Coordinates (X, Y, Z), ($X_1$, $Y_1$, $Z_1$) and ($X_2$, $Y_2$, $Z_2$) denote lab coordinate, camera 1 coordinate and camera 2 coordinate, respectively. $\varphi_1$ and $\varphi_2$ denote the angle between $X_1$ and X direction, $X_2$ and X direction, respectively. $\phi = \varphi_2 - \varphi_1$ denotes the angular separation of two projections in the XMPI setup, which is 47.7 degrees in the experiments conducted in this work. Note that the cameras can only provide 2D images, so $X_1$ and $X_2$ coordinates cannot be directly read from the acquired images.

Suppose we have a particle located at $r = (x, y)$ in XY plane. From two cameras, $y_1$ and $y_2$ can be read to determine the particle position, as it is obvious from Fig. 7(a) that $y_1 = R - d_1$ and .$y_2 = R - d_2$. Describing the particle position in two camera coordinates $(x_1, y_1)$ and $(x_2, y_2)$, respectively, we have:

$$\begin{cases} x = x_1 \cos\varphi_1 - y_1 \sin\varphi_1 = x_2 \cos\varphi_2 - y_2 \sin\varphi_2 \\ y = x_1 \sin\varphi_1 + y_1 \cos\varphi_1 = x_2 \sin\varphi_2 + y_2 \cos\varphi_2 \end{cases} \tag{2}$$

After eliminating $x_1$ and $x_2$, which cannot be directly read from the images, we have:

$$\begin{cases} x = \dfrac{y_1 \sin\varphi_1 + y_1 \sin(\varphi_1 - 2\varphi_2) + y_2 \sin\varphi_2 - y_2 \sin(2\varphi_1 - \varphi_2)}{\cos(2\varphi_1 - 2\varphi_2) - 1} \\ y = \dfrac{2(y_1 \sin\varphi_2 - y_2 \sin\varphi_1)\sin(\varphi_1 - \varphi_2)}{\cos(2\varphi_1 - 2\varphi_2) - 1} \end{cases} \tag{3}$$

In equation 3, the uncertainty of both $x$ and $y$ originate from the uncertainty of $y_1$ and $y_2$. Therefore, we have:

$$\begin{cases} \sigma_x = \sqrt{\left(\dfrac{\partial x}{\partial y_1} \cdot \sigma_{y_1}\right)^2 + \left(\dfrac{\partial x}{\partial y_2} \cdot \sigma_{y_2}\right)^2} \\ \sigma_y = \sqrt{\left(\dfrac{\partial y}{\partial y_1} \cdot \sigma_{y_1}\right)^2 + \left(\dfrac{\partial y}{\partial y_2} \cdot \sigma_{y_2}\right)^2} \end{cases} \tag{4}$$

Furthermore, for calculating the uncertainty of the radial position in XY plane, we have:

$$\begin{cases} r = \sqrt{x^2 + y^2} \\ \sigma_r = \dfrac{\sqrt{\sigma_x^2 x^2 + \sigma_y^2 y^2}}{r} \end{cases} \tag{5}$$

In PTV conducted in this work, we focus on getting the radial position of each particle, regardless of the orientation of the lab coordinate in XY plane (defined by $\varphi_1$). Therefore, for simplicity in estimating $\sigma_r$, we select $\varphi_1 = (90° - \phi)/2$ so that $\sigma_x = \sigma_y = \sigma_r/\sqrt{2}$.

Given that both cameras are identical and independent regarding their detection efficiency and spatiotemporal resolution despite their slightly different sample-detector distances, the uncertainties of $y_1$ and $y_2$ are the same. Given that the cameras have the same resolution in both axes, we further assume that such uncertainty is the same as $\sigma_z$, which describes the uncertainty of vertical position in the lab coordinate. We estimate $\sigma_z$ by considering three main sources of error. The first is related to the instability of the beamlet, giving $\sigma_1 = 1$ pix. The second is related to the particle segmentation as described in step 2 of the MyPTV analysis. Considering the difficulty of deciding the center of each particle, especially when several particles contact or interact with each other, a reasonable uncertainty of $\sigma_2 = 2$ pix is used. The third comes from the camera calibration process, resulting in $\sigma_3 = 1$ pix. Therefore, we have:

$$\sigma_{y_1} = \sigma_{y_2} = \sigma_z = \sqrt{\sigma_1^2 + \sigma_2^2 + \sigma_3^2} = 2.45 \text{ pix} \tag{6}$$

Using equations (3)-(6), the relationship of $\phi$ and the ratio $\sigma_r/\sigma_z$ is shown in Fig. 7(b). The uncertainty in radial position decreases when angular separation $\phi$ increases towards $90°$. In this work, $\phi = 47.7°$, so the uncertainties of the position of the tracked particle are: $\sigma_z = 2.45 \text{ pix}$;
$\sigma_r = \sqrt{2}\sigma_x = \sqrt{2}\sigma_y = 1.91\, \sigma_z = 4.7 \text{ pix}$.

According to step 6 of the MyPTV analysis, linear fitting of the trajectory is implemented to calculate the velocity. Suppose we have multiple points $(z_i, t_i)$ with $i = 1, 2, \dots, n$ for a certain trajectory, where $n$ indicates the total number of time points per trajectory; $z_i$ is the vertical position given by the tracking results at time $t = t_i$. The vertical velocity $u_z$ is the slope of the fitted line with an uncertainty $\sigma_{u_z}$ given in equation (7), which is dependent on both $\sigma_z$ and the temporal length of the trajectory (the number of frames per trajectory) as is shown in Fig. 7(c). In the experiment presented in Fig. 3(a)-(c), the average Pearson correlation coefficient for all 2471 tracked particles is 0.9992, indicating that the acceleration of particles in $z$ direction is negligible.

$$\sigma_{u_z} = \sigma_z \cdot \sqrt{\frac{1}{\sum (t_i - \bar{t})^2}} \tag{7}$$

To avoid underestimating the error in calculating the vertical velocity, we use the trajectory with minimal temporal length to estimate $\sigma_{u_z}$ for all the tracked particles in each experiment. For example, the minimal trajectory length is 20 frames in the experiment presented in Fig. 3(a)-(c), resulting in $\sigma_{u_z} = 0.005 \text{ mm/s}$.

To sum up, in this work, we can estimate that the vertical position uncertainty $\sigma_z = 2.45 \text{ pix} = 3.2\ \mu\text{m}$; the radial position uncertainty is $\sigma_r = 4.7 \text{ pix} = 6.1\ \mu\text{m}$. The uncertainty of vertical velocity depends on the temporal length of the trajectory. A longer trajectory results in lower uncertainty.

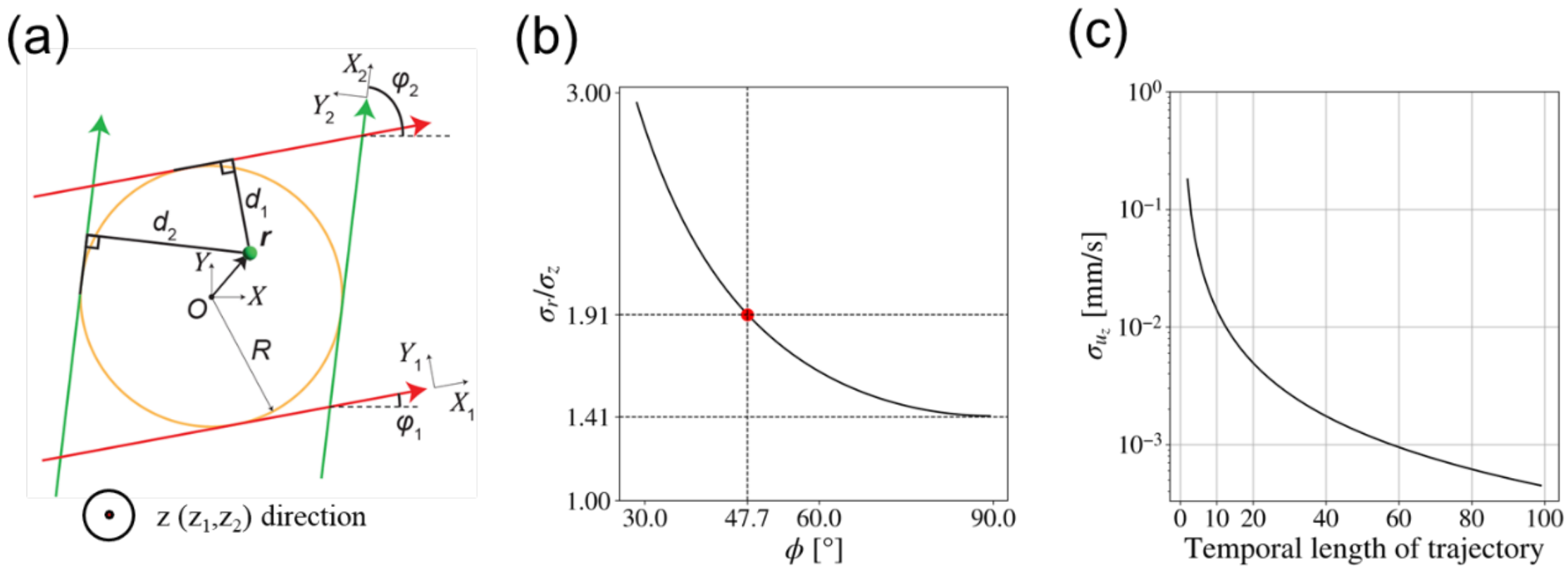

**Fig. 7** (a) Coordinate system used for error analysis; (b) Relationship between the XMPI angular separation $\phi$ and the radial position uncertainty described in ratio of $\sigma_r/\sigma_z$; (c) Relationship between the temporal length of a specific trajectory and its uncertainty of vertical velocity.

## Appendix E Microparticle tracking at different particle densities and flow rates

We implemented microparticle tracking for various particle densities and flow rates, using the same workflow as shown in the Methods section of the main manuscript. Fig. 8 shows the results at the same flow rate setting of 0.1 mL/h in three different particle densities (0.05 wt.%, 0.1 wt.%, 0.2 wt.%), respectively. The first row (a)-(c) shows the particle velocity distribution in different horizontal positions, and the second row (d)-(f) shows how the particles follow the expected theoretical velocity profile of an incompressible laminar flow through a cylinder (solid curve). A $3\sigma$ error band is created based on the uncertainties $\sigma_r$ and $\sigma_{u_z}$ calculated in Appendix D. The proportions of tracked particles within the error band are 97.3%, 95.8%, and 90.0%, respectively.

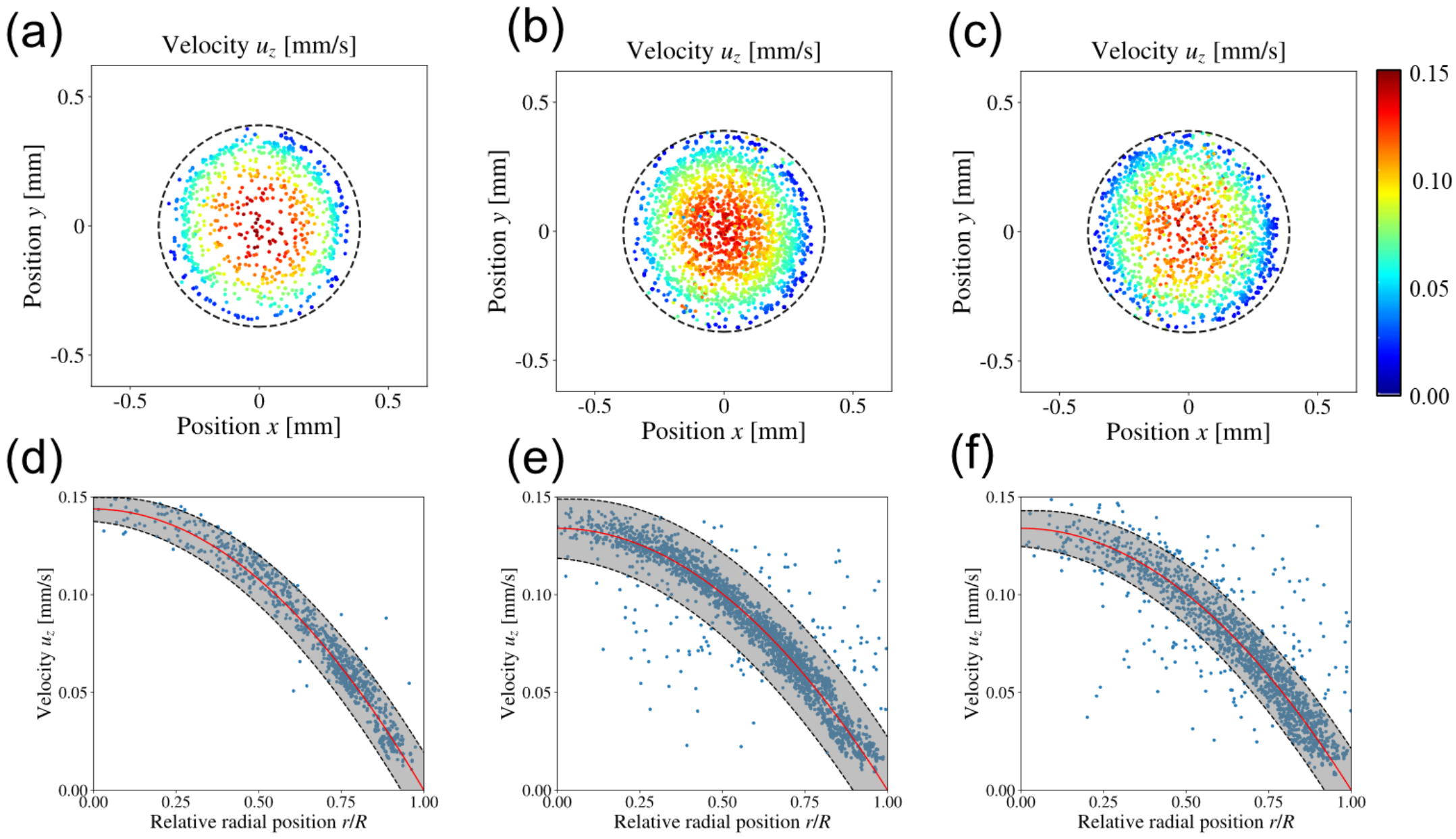

**Fig. 8** Microparticle tracking at a flow rate setting of 0.1 mL/h; the particle densities are (a) 0.05 wt.%, (b) 0.1 wt.%, and (c) 0.2 wt.%. The lower figures (d)-(f) show the particle velocities as function of radial positions at the same concentrations, where the shadowed region indicates the $3\sigma$ error band. The proportions of tracked particles within the error band are 97.3%, 95.8%, and 90.0%, respectively.

Analogously, Fig. 9 shows the results at the same particle density of 0.05 wt.% at three different flow rate settings (0.1 mL/h, 0.2 mL/h, 0.5 mL/h), respectively. The first row (a)-(c) shows the particle velocity distribution in different horizontal positions and the second row (d)-(f) shows how the particles follow the expected theoretical velocity profile of an incompressible laminar flow through a cylinder (solid curve). The proportions of tracked particles within the error band are 97.3%, 95.7%, and 96.3%, respectively.

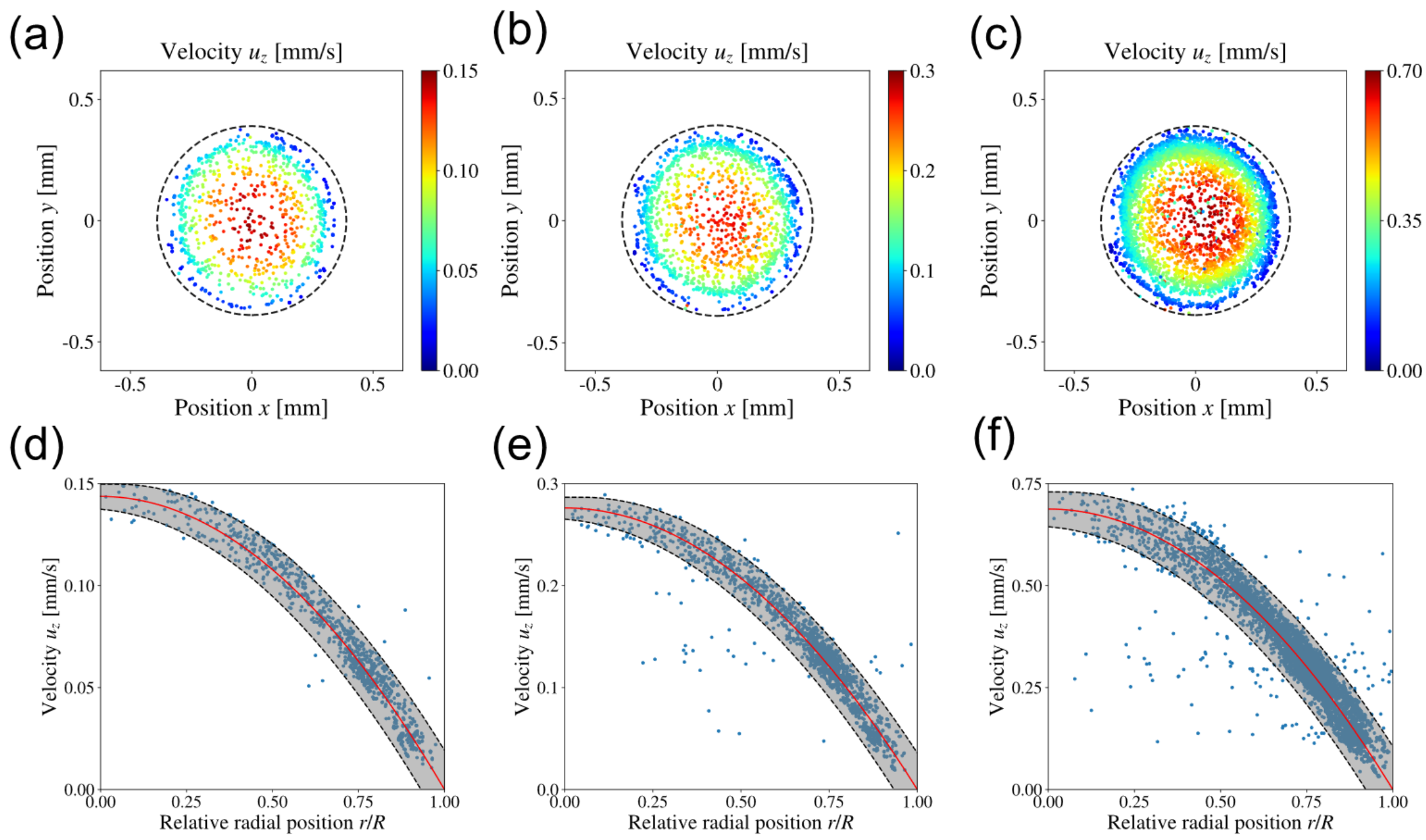

**Fig. 9** Microparticle tracking at a particle density of 0.05 wt.% at three different flow rate settings (a) 0.1 mL/h, (b) 0.2 mL/h, and (c) 0.5 mL/h. The lower figures (d)-(f) show the particle velocities as function of radial positions at the same concentrations, where the shadowed region indicates the $3\sigma$ error band. The proportions of tracked particles within the error band are 97.3%, 95.7%, and 96.3%, respectively.

## Appendix F Optical flow (OF) analysis of dense particle suspensions

Optical flow (OF) was performed on a 5-second sequence of flat-field corrected images (200 images) of SHGS at 10 wt.% using the RAFT deep learning algorithm opticalFlowRAFT in MATLAB R2024b (MathWorks 2024). The results are visible in Fig. 10. Note that flow velocity vectors from the OF calculation are assigned to every pixel in the flat-field corrected projections, but for clarity, only every 20th vector in horizontal and vertical directions are shown. The velocities were scaled using the known pixel size and frame rate, then averaged over time and vertical direction.

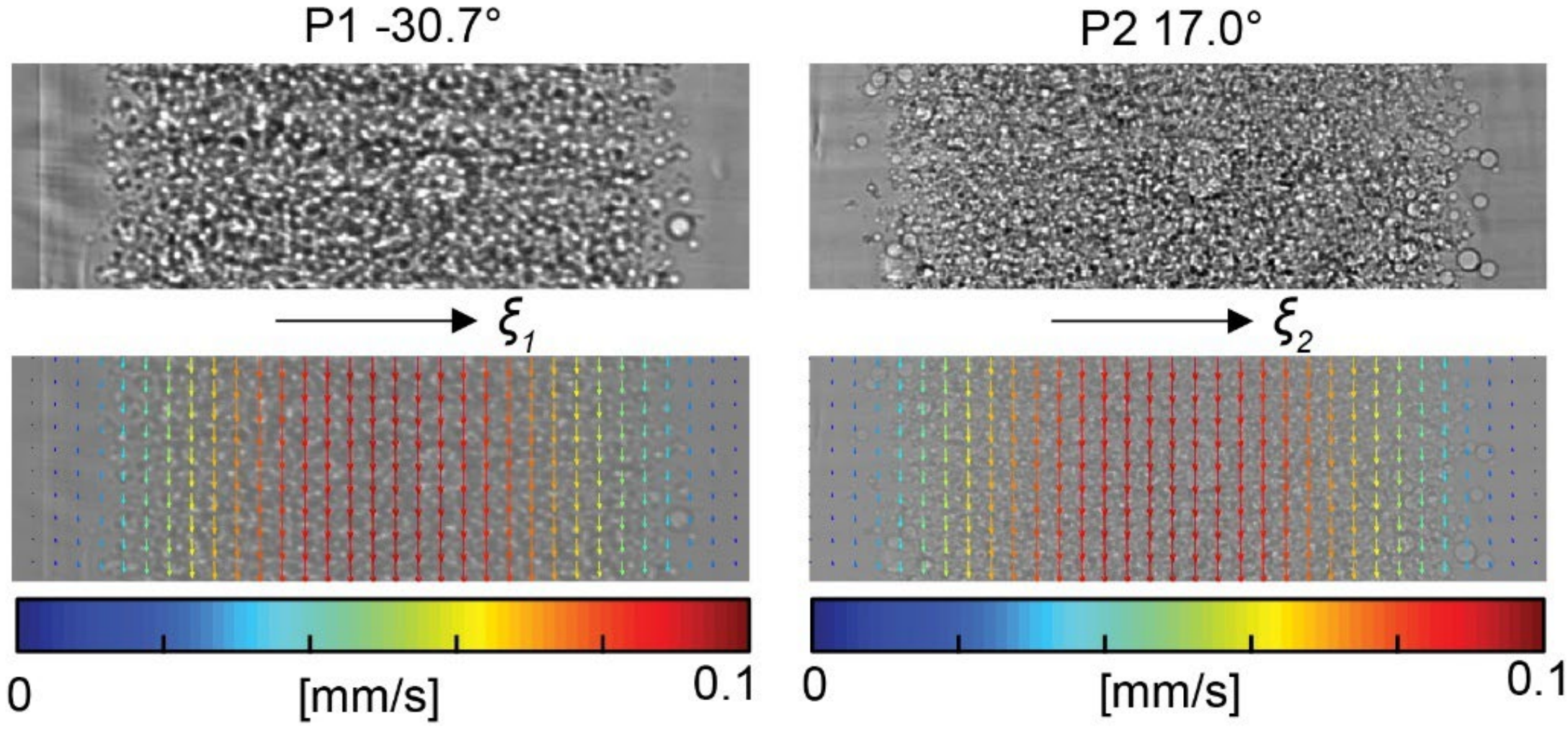

**Fig. 10** Vector fields of the projected velocities resulting from the Optical Flow (OF) analysis. Although a velocity vector is assigned for every pixel, only every 20th vector in horizontal and vertical directions are shown for clarity.

Projected wall positions were determined manually from the raw images where they are clearly visible (see Fig. 11). In order to avoid erroneous vectors from image noise, we define a measurement window 50 pixels away from the walls where there are particles present throughout the image height.

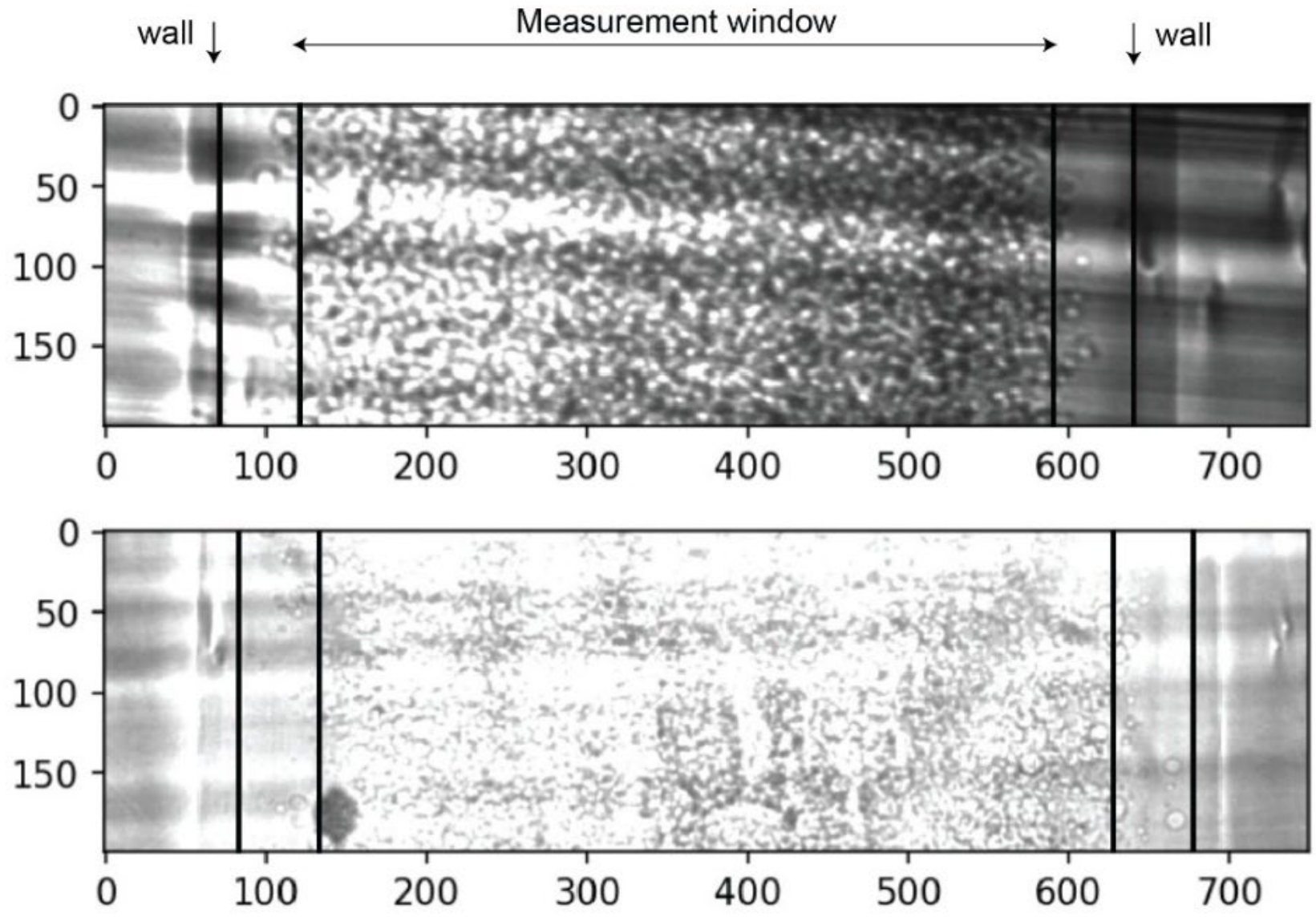

**Fig. 11** Projected wall positions are determined from the raw images before flat-field correction (P1 upper, P2 lower) and a measurement window is determined 50 pixels from the projected walls.

After averaging over time and vertical direction we arrive at the average projected velocity in Fig. 12. The two projected curves look very similar from the two projections reflecting the cylindrical symmetry of the flow. It is important to note that these curves are integrated mean velocities along the line of sight, which in the center region contains both slow-moving particles close to the walls and fast-moving particles close to the center.

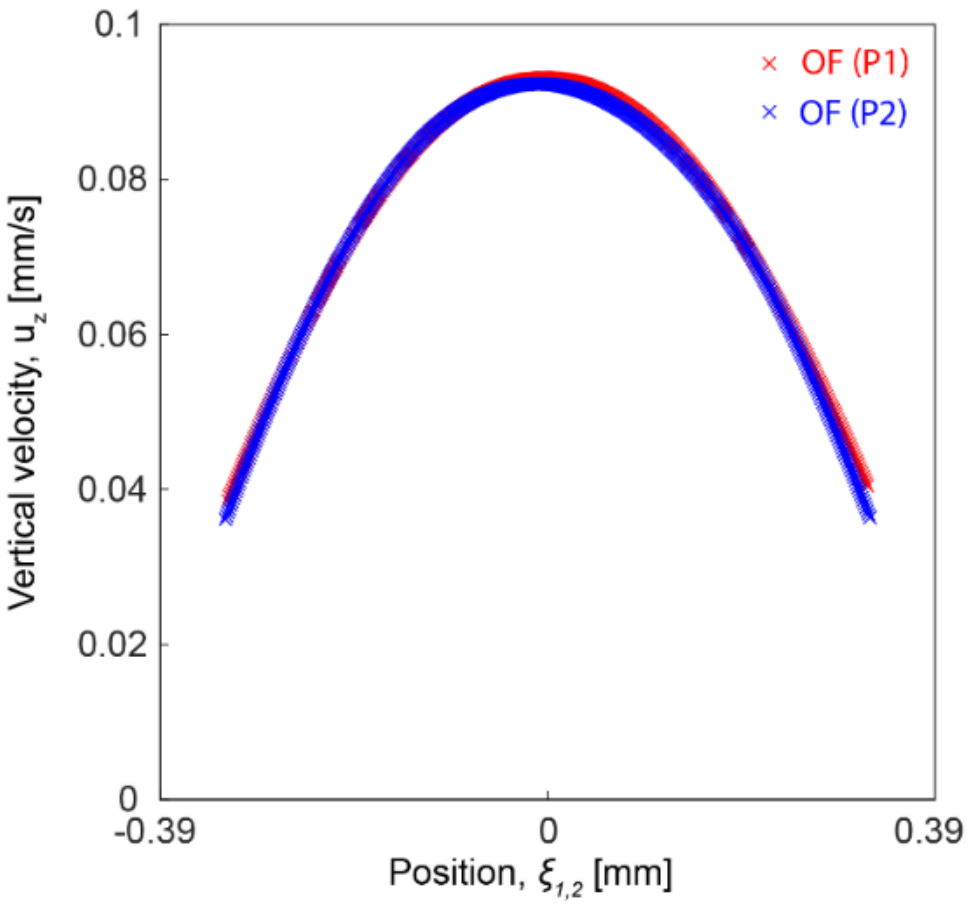

**Fig. 12** Projected velocity profiles from the OF analysis.

At this particle concentration (10 wt% corresponding to 9 vol%), suspension viscosity can be assumed to be shear-rate independent, and the radial velocity profile should be comparable with the velocity profile from 3D Particle Tracking Velocimetry (3D-PTV) velocity profile of dilute particles given the same flow rate.

To compare the OF results from the dense suspension with the 3D-PTV velocity profile, the latter is firstly projected using a Monte Carlo methodology. A total of $10^7$ artificial particles were randomly sampled within a circle with a diameter of 2R (R = 0.39 mm) and assigned a velocity depending on its radial position according to the fitted data of Q and R from the 3D-PTV results according to $u_z = (2Q/(\pi R^2))(1\text{-}r^2/R^2)$. This is seen in Fig. 13(a) using a reduced number of artificial particles for clarity.

The particles are then subsequently binned in one direction (x) while averaged in the other (y), see Fig. 13(b) (binning is coarser in figure for clarity), to generate the projected velocity profile $u_z(\xi)$ that can be compared with the OF results, as seen in Fig. 13(c). In this figure, it can be seen that the measured velocity profile is seemingly higher than the estimation from the dilute system. However, as is seen in the experimental results, there is a particle-free layer closest to the wall. This has a big effect on the projected velocity as the density of particles is not uniform over the cross section. With a lack of slow-moving wall-particles, the center velocity will be perceived higher than the projected 3D-PTV profile.

To test the influence of a particle-free layer of thickness $\Delta r$, the same Monte Carlo sampling is done, assuming a uniform density of artificial particles within a region of $r < R - \Delta r$. At $R - \Delta r < r < R$, the density is zero and no sampling of particles, illustrated in Fig. 13(d).

By comparing the projected velocity profile given a few values of $\Delta r$, the center velocity has a good agreement at a value of $\Delta r$ = 0.02 mm, as seen in Fig. 13(e).

Although improved, there is not a perfect match of the shape of the velocity profile, which likely arises from the radial concentration of particles not being well represented with a step function. To further improve the results, a measurement without phase contrast edge enhancement could lead to an attenuation-based projected concentration profile. Complementing the mean projected velocity measurements with the projected concentration profile, assuming cylindrical symmetry and no swirl, the radial concentration profile and 3D velocity profile can be recovered via an Abel transform, which can be done in a single projection.

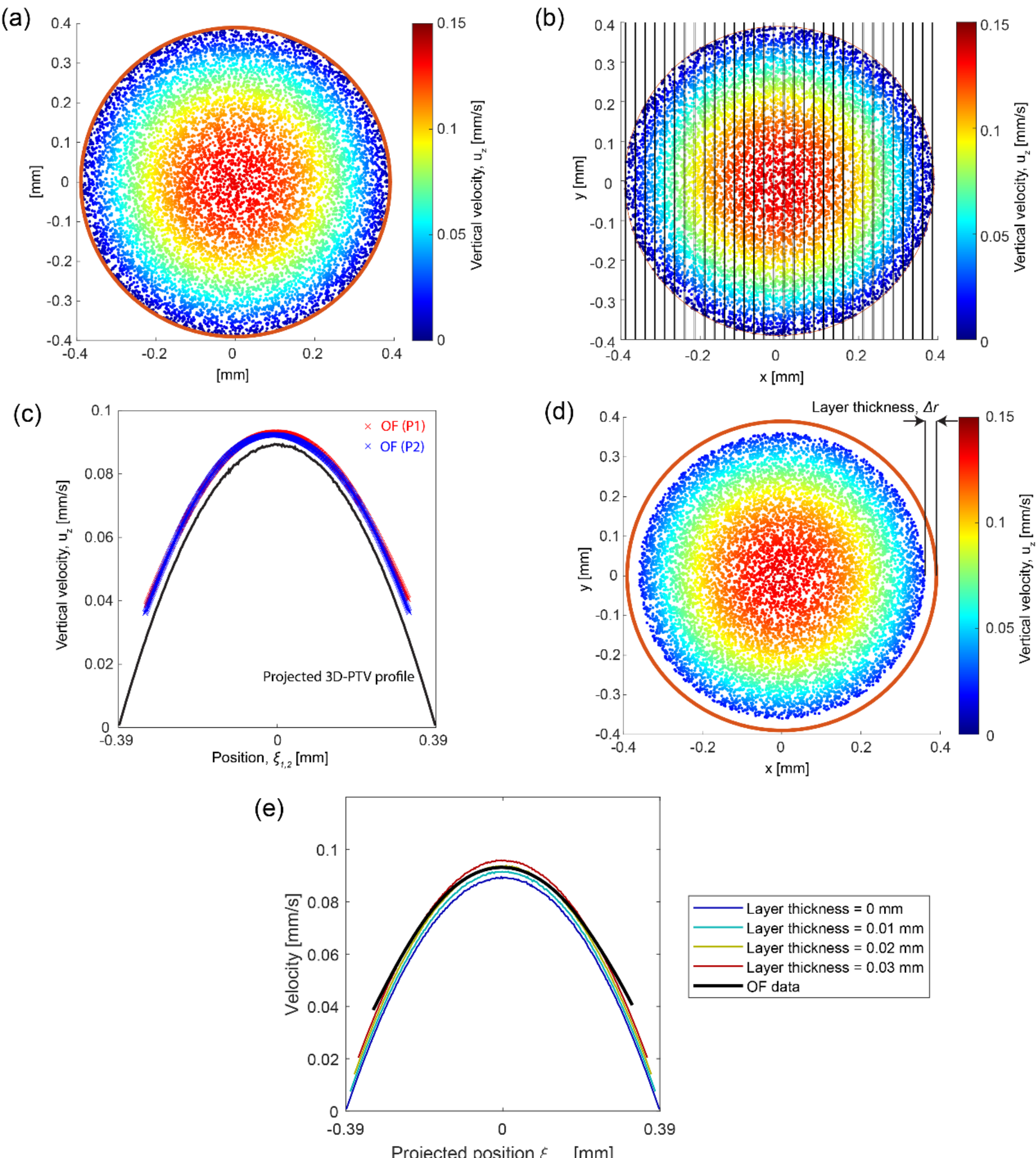

**Fig. 13** Comparison of 3D-PTV velocity profile in a dilute suspension with the projected velocity profiles from the OF analysis of P1. (a) Artificial particles are randomly sampled within the tube diameter with velocities assigned from their radial position based on the velocity profile from the 3D-PTV experiment. (b) Particles are binned according to their projected position and averaged in each bin to get a projected velocity that is comparable with the OF results. (c) The projected 3D-PTV profile compared to the OF results, where the underestimation likely arises from a non-uniform radial concentration distribution. (d) Artificial particles randomly sampled while assuming a particle-free layer thickness Δr. (e) Influence of Δr on the projected 3D-PTV profile, where a Δr ≈ 0.02 mm leads to the best agreement with OF velocity profile.

Ultimately, having two views further allows us to assess *e.g.* cylindrical symmetry and quantify projected flow fields without this symmetry. Even though there is not enough information to directly find the flow field through *e.g.* Abel inversion, the demonstration here shows how projected profiles can be matched between simulations and experiments. By testing various scenarios and parameters with a CFD model and matching projected profiles, the simulation can be used as a digital twin to illustrate most likely 3D velocity fields that would generate the projected velocity and concentration profiles in the experiment.

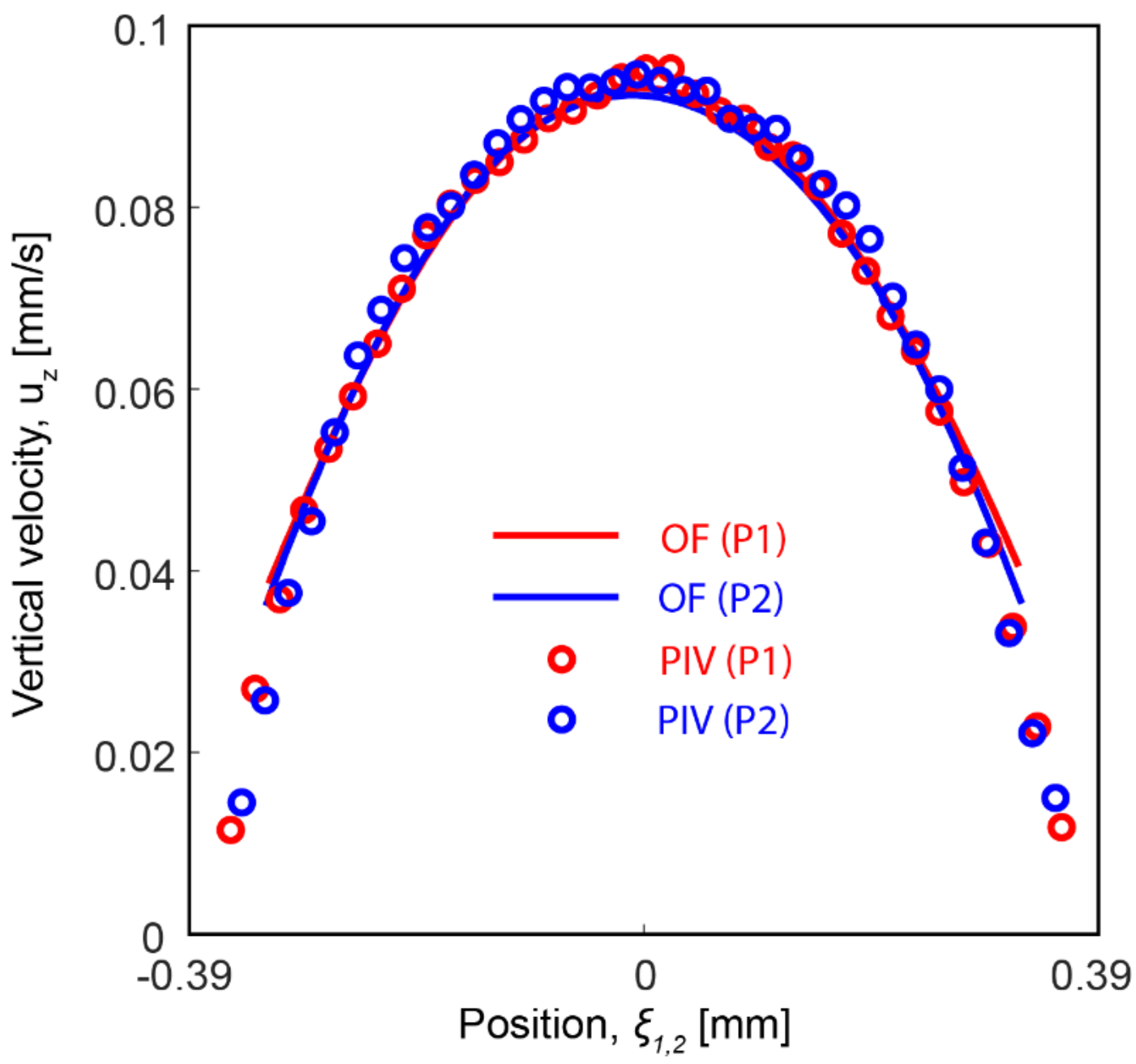

**Fig. 14** Comparison of OF analysis and Particle Image Velocimetry (PIV).

As a final remark, the analysis of the projected flow of dense suspensions can be analyzed also with particle image velocimetry (PIV) methods. As a comparison, PIV was performed on the same data using PIVlab v3.02 (Thielicke and Sonntag 2021) in MATLAB R2023b. The PIV algorithm used FFT window deformation with one pass using an interrogation area of 32 pixels and a step size of 16 pixels. The results are compared with the OF results in Fig. 14. Although the best practice of PIV should be to not use large tracers as done in this work, we find that the results with the two methodologies are almost identical.

**Acknowledgments**

Financial support from Swedish Research Council (Starting Grant 2022-02863) is gratefully acknowledged. This work has received funding from ERC-2020-STG 3DX-FLASH 948426. We acknowledge MAX IV Laboratory for time on Beamline ForMAX under Proposal 20231192. Research conducted at MAX IV, a Swedish national user facility, is supported by the Swedish Research Council under contract 2018-07152, the Swedish Governmental Agency for Innovation Systems under contract 2018-04969, and Formas under contract 2019-02496.

**Data availability**

Data will be made available on request.

**CRediT authorship contribution statement**

**Tomas Rosén:** Conceptualization, Methodology, Formal analysis, Funding acquisition, Resources, Investigation, Visualization, Project administration, Supervision, Writing – original draft. **Zisheng Yao:** Methodology, Formal analysis, Resources, Investigation, Visualization, Writing – original draft. **Jonas Tejbo:** Resources, Investigation, Writing – review & editing. **Patrick Wegele:** Investigation, Writing – review & editing. **Julia K. Rogalinski:** Methodology, Resources, Investigation, Writing – review & editing. **Frida Nilsson:** Conceptualization, Resources, Writing – review & editing. **Kannara Mom:** Methodology, Resources, Investigation, Writing – review & editing. **Zhe Hu:** Methodology, Resources, Investigation, Writing – review & editing. **Samuel A. McDonald:** Resources, Investigation, Writing – review & editing. **Kim Nygård:** Resources, Investigation, Writing – review & editing. **Andrea Mazzolari:** Methodology, Resources, Writing – review & editing. **Alexander Groetsch:** Investigation, Writing – review & editing. **Korneliya Gordeyeva:** Resources, Investigation, Writing – review & editing. **L. Daniel Söderberg:** Conceptualization, Funding acquisition, Writing – review & editing. **Fredrik Lundell:** Conceptualization, Writing – review & editing. **Lisa Prahl Wittberg:** Conceptualization, Resources, Investigation, Writing – review & editing. **Eleni Myrto Asimakopoulou:** Conceptualization, Methodology, Resources, Investigation, Writing – review & editing. **Pablo Villanueva-Perez:** Conceptualization, Methodology, Funding acquisition, Resources, Investigation, Visualization, Project administration, Supervision, Writing – original draft.

**Declaration of competing interest**

The authors declare that they have no competing interests.

**Declaration of generative AI and AI-assisted technologies in the writing process.**

During the preparation of this work the author(s) used ChatGPT 4o and Microsoft CoPilot in order to assist with grammar, spelling, and stylistic improvements. After using this tool/service, the author(s) reviewed and edited the content as needed and take(s) full responsibility for the content of the published article.